\documentclass[cup7b]{cupbook}
\usepackage{graphicx}
\usepackage{amssymb,amsmath,epsfig}
\newcommand{\beq}{\begin{equation}}
\newcommand{\eeq}{\end{equation}}
\newcommand{\barr}{\begin{array}}
\newcommand{\earr}{\end{array}}

\newcommand{\Hi}{\mathcal{H}}
\newcommand{\Ai}{\mathcal{A}}
\newcommand{\m}{\Phi}

\newcommand{\cs}{C${}^*$}
\def\one{{\mathchoice{\rm 1\mskip-4mu l}{\rm 1\mskip-4mu l}{\rm 1\mskip-4.5mu l}{\rm
1\mskip-5mu l}}}
\def\ket#1{| #1 \rangle}

\def\kb#1#2{|#1\rangle\!\langle #2 |}

\def\A{{\cal A}}
\def\B{{\cal B}}

\def\H{{\cal H}}
\def\K{{\cal K}}

\def\P{{\mathcal P}}

\def\S{{\cal S}}
\def\U{{\cal U}}

\def\Tr{{\mathrm{Tr}}}

\newcommand{\qforal}{\quad\text{for all}\quad}

\newtheorem{definition}{Definition}

\newtheorem{thm}{Theorem}

\newcommand{\G}{\Gamma}

\newcommand{\Aevol}{{\cal A}_{\mbox{\footnotesize{ evol}}}}

\begin{document}
\author[F. Markopoulou]{Fotini Markopoulou\\
Perimeter Institute for Theoretical Physics}

\chapter{Towards Gravity from the Quantum}

\section{Introduction}

The different approaches to quantum gravity, almost all of which are described in detail in the rest of this volume,  can be classified according to what they say about spacetime and gravity.  

First, there are the quantum field theory-like approaches (string theory, black hole thermodynamics, etc.).  It is natural to push quantum field theory to its limits since it is our best theory so far.  Results coming from traditional quantum field theory on curved spacetime are clearly limited since the spacetime is not dynamical.  String theory obtains the graviton as an effective excitation, but the fundamental quantities of the theory in its current formulation refer to a fixed background.  There are
 indications that there may be indirect ways in which a dynamical spacetime and gravity is present in the theory.     It is non-trivial to come to a conclusive verdict but the strategy is that general relativity should arise as an effective theory.  

The so-called background independent approaches to quantum gravity state that the fundamental quantum theory ought to explicitly possess a dynamical spacetime.  Presumably because all these originate in general relativity,  spacetime is expected to be fundamental except quantum, in the sense that there are geometric and gravitational degrees of freedom in the fundamental theory.  The implementation is either via a strict quantum version of classical gravity (straight quantization as in Loop Quantum Gravity) or a more general quantum superposition of spacetimes (as in traditional causal sets or Causal Dynamical Triangulations). 
 Morally, such approaches stay close to general relativity, in practice, there are difficult issues related to  dynamics, observables and, often, an ill-defined path integral.  
 An important recent advance is the new approach of Causal Dynamical Triangulations \cite{CDT}, a sum-over-geometries theory.  
In CDT  time is treated fundamentally different than the other approaches.  Strong causality constraints  are crucial to the good properties of the theory, indicating that some further physical input than a  plain quantum version of general relativity may be  necessary.

Then there are the
condensed-matter approaches to quantum gravity (such as \cite{CondMatt,Dre}).    These also explore the idea that gravity may be an emergent theory.   Literature here ranges from just an analogy between spacetime thermodynamics and condensed matter systems to radical ideas of a pre-spacetime condensed  matter system (the subtleties of these are discussed by Dreyer in this volume).   

It is peculiar that the approaches that advocate that gravity is only an effective theory (string theory, condensed matter) are based explicitly on a spacetime being present while approaches that are background independent consider gravity to be fundamental. 

Here, we will
advocate an approach orthogonal to the quantum field theory-like approaches above (we are background independent) but also orthogonal to the usual background independent approaches (there will be no fundamental degrees of freedom for the gravitational field).  That is, we will work with a microscopic theory that  is a pre-spacetime quantum theory. The macroscopic description of this is in terms of dynamically selected excitations, that is, coherent degrees of freedom that survive the microscopic evolution to dominate our scales.   We propose that 
it is properties of the interactions of these excitations that we understand as spacetime.  Furthermore, as does Dreyer in this volume, we will conjecture that the background independence will lead to a dynamical spacetime.   

The mathematical formalism we will base the discussion on is that of a quantum causal history (QCH) \cite{MarQCH,HawMarSah}.  This is a locally finite directed network of finite-dimensional quantum systems (section \ref{sectionQCH}).  The requirement of local finiteness is a simple implementation of the expectation that there really are only a finite number of degrees of freedom in a finite volume (arguments for which are well-known and we have reviewed them elsewhere \cite{MarWhe}).  
QCHs have had many lives of distinct physical interpretations:  1) A discrete quantum field theory with varying number of degrees of freedom in time (section \ref{sectionQFT}).  2)  A causal quantum geometry in the traditional quantum sum-over-histories setup (section \ref{sectionQSF}).   Such theories encounter difficult issues when it comes to their low energy limit (section \ref{sectionLEP}) 3)  A quantum information processor which can be used as a pre-spacetime theory (section \ref{sectionQIP}).

In section \ref{sectionNS}, we will take up the idea that the effective description of a background independent theory can be characterized by the dynamics of coherent excitations in the fundamental theory and implement it by importing the method of noiseless subsystems from quantum information theory.   We give an example of such coherent excitations in a common kind of graph-based theory in section \ref{sectionCohDofsQGeo}.  

It is really section \ref{sectionQIP} that is the central one of this paper.  It contains a departure of the traditional implementations of background independence to the simple one that is achieved by a pre-spacetime theory.  The language of quantum information theory is used here, not due to a philosophical preference for information (as in ``it from bit'') but simply because unnecessary references to a background spacetime can be eliminated.  In section \ref{sectionSymmetry}, we implement the idea that all we can mean by a Minkowski spacetime is that all coherent degrees of freedom and their interactions 
  are Poincar\'{e} invariant at the relevant scale by an appropriate extension of the notion of a noiseless subsystem.  
  
A spacetime emergent from a background independent theory raises issues regarding emergent locality vs fundamental locality (section \ref{sectionlocality}). We are optimistic that this is not a problem but, rather, an opportunity for quantum gravity phenomenology that is not tied to the Planck scale.  It also raises questions about the role of time which we discuss in section \ref{sectionTime}.  Finally, in the speculative section \ref{sectionConjecture}, we conjecture that this direction may provide a natural place and explanation for the Einstein equations.   We summarize in the Conclusions.

\section{Quantum Causal Histories}
\label{sectionQCH}
A quantum causal history is a locally finite directed network of finite-dimensional quantum systems 
We start with the network, a directed graph.  In sections \ref{sectionQFT}, \ref{sectionQSF} and \ref{sectionQIP},  the graph will acquire meaning as the regions of a classical spacetime, a causal set and a pre-spacetime circuit of quantum operations respectively, each with distinct implications for their physical content.  

\subsection{The directed acyclic graph}\label{sectionGamma}

Let $\Gamma$ be a directed graph with vertices $x\in V(\Gamma)$
and directed edges $e\in E(\Gamma)$. The {\it source} $s(e)$  and
{\it range} $r(e)$ of an edge $e$ are, respectively, the initial
and final vertices of $e$. A (finite) path $w=e_k \cdots e_1$ in
$\Gamma$ is a sequence of edges of $\Gamma$ such that $r(e_i) =
s(e_{i+1})$ for $1\leq i < k$. If $s(w) = r(w)$ then we say $w$ is
a {\it cycle}. We require that $\Gamma$ has no
cycles\footnote{This condition was initially
motivated by $\Gamma$ being a causal set \cite{CauSet}, in which case, a cycle is a closed timelike loop (see section \ref{sectionQSF}).  The same condition is also natural if the quantum causal history is a
quantum computer with $\Gamma$ the circuit.}. 

If there exists a
path $w$ such that   $s(w)=x$ and $r(w)=y$
 let us write $x\leq y$ for
the associated partial ordering.  We call such vertices {\em related}.  Otherwise, they are {\em unrelated} and we use $x\sim y$ to denote this.  Given any $x\leq y$, we require that there are finitely many $z\in V(\Gamma)$
such that $x\leq z \leq y$.  This is the condition of {\em local finiteness}.

We now wish to associate quantum systems to the graph.  They will be related by the appropriate maps, {\em completely positive maps}, the basics of which we review next.  

\subsection{Completely Positive Maps}
\label{sectionCP}

Completely positive maps, or {\em quantum channels}, are commonly used to describe evolution of open quantum systems (see, for example, \cite{NieChu}).  

Let $\Hi_S$ be the state space of a quantum system in contact with
an environment $\Hi_E$. The standard characterization of evolution
in open quantum systems starts with an initial state in the system
space that, together with the state of the environment, undergoes
a unitary evolution determined by a Hamiltonian on the composite
Hilbert space $\Hi = \Hi_S \otimes \Hi_E$, and this is followed by
tracing out the environment to obtain the final state of the
system. 

The associated evolution map, or ``superoperator'',
$\Phi:{\cal A}(\Hi_S)\rightarrow{\cal A}(\Hi_S)$ between the corresponding matrix algebras of operators on the respective Hilbert spaces is necessarily completely
positive (see below) and trace preserving. More generally, the map
can have different domain and range Hilbert spaces. Hence the
operational definition of a {\it quantum channel} (or quantum
evolution, or quantum operation) from a Hilbert space $\Hi_1$ to
$\Hi_2$, is a completely positive, trace preserving map $\Phi:
{\cal A}(\Hi_1) \rightarrow {\cal A}(\Hi_2)$.  In more detail:

\begin{definition}
A {\it completely positive} (CP) map  $\Phi$ is a linear map $\Phi
: {\cal A}(\Hi_1)\rightarrow{\cal A}(\Hi_2)$ such that the maps
\[
 id_k \otimes \Phi :  M_k \otimes {\cal A}(\Hi_1) \rightarrow
 M_k \otimes {\cal A}(\Hi_2)
\]
are positive for all $k\geq 1$. 
\end{definition}
Here we have written $M_k$ for
the algebra ${\cal A}(\mathbb{C}^k)$ represented as the $k\times k$
matrices with respect to a given orthonormal basis. (The CP
condition is independent of the basis that is used.)

Complete positivity (as opposed to positivity) is needed for such maps to describe evolution of physical systems.  Examples of positive operators  are known that do not describe physical evolution.  Although there are examples of non-CP evolution, the CP case is generic enough that little is lost if we remain with it for the present work and take advantage of the relevant powerful properties of CP operators.

A fundamental technical device in the study of CP maps is the {\it
operator-sum representation} theorem of Choi and Kraus. 
We will use this in extracting effective degrees of freedom from a quantum causal history. 

\begin{thm}{\bf\rm(Choi and Kraus)}
For every CP map $\Phi$ there
is a set of operators $\{E_a\}\subseteq {\cal A}(\Hi_1,\Hi_2)$ such that
\begin{eqnarray}\label{opsum}
\Phi(\rho) &=& \sum_a E_a \rho E_a^\dagger \qforal
\rho\in{\cal A}(\Hi_1).
\end{eqnarray}
\end{thm}
We shall write $\Phi = \{E_a\}$ when the $E_a$ satisfy
Eq.~(\ref{opsum}) for $\Phi$. The family $\{E_a\}$ may be chosen
with cardinality $|\{E_a\}|\leq \dim(\Hi_1)\dim(\Hi_2)$, and is
easily seen to be non-unique\footnote{However, if $\{E_a\}$ and
$\{F_b\}$ are two families of operators that implement the same
channel $\Phi$, then there is a scalar matrix $U=(u_{ab})$ such
that $E_a = \sum_b u_{ab} F_b$ for all $a$.}.

The class of CP maps that are quantum channels satisfy an extra
constraint. Specifically, note that when $\Phi$ is represented as
in (\ref{opsum}), trace preservation is equivalent to the identity
\begin{eqnarray}\label{tracepreserve}
 \sum_a E_a^\dagger  E_a = \one_{\Hi_{1}}.
\end{eqnarray}
Thus, a quantum channel $\Phi$ is a map which satisfies
(\ref{opsum}) and (\ref{tracepreserve}) for some set of operators
$\{E_a\}$.

\subsection{Construction and definition of a quantum causal history}

The construction of a  quantum causal history \cite{MarQCH} starts with a  directed graph $\Gamma$
and assigns to every
vertex $x\in V(\Gamma)$ a finite-dimensional Hilbert space
$\Hi(x)$  and/or a matrix algebra $\Ai(\Hi(x))$ (or $\Ai(x)$ for short) of operators acting on $\Hi(x)$.  It is best to regard the algebras as the primary objects \cite{HawMarSah}, but we will not make this distinction here.  
For every edge $e\in E(\Gamma)$ there is a quantum channel
\beq
\Phi_e: {\cal A}(s(e))\rightarrow{\cal A}(r(e)),\label{eq:qchphi}
\eeq
where ${\cal A}(x)$ is the full matrix algebra on $\Hi(x)$. \

A {\em parallel set} $\xi\subseteq E(\Gamma)$ is defined by the
property that $x\sim y$ whenever $x,y\in\xi$. The algebra ${\cal
A}(\xi) = \otimes_{x\in\xi} {\cal A}(x)$ acts on the composite
system Hilbert space ${\cal H}(\xi) = \otimes_{x\in\xi} {\cal
H}(x)$. A parallel set $\xi$ is a {\em complete source} of $x$ if all paths $w$ with $r(w)\equiv x$ have $s(w)\in\xi$.  Conversely, a parallel set $\zeta$ is a {\em complete range} of $x$ if all paths $w$ with source $s(w)\equiv x$ have range in $\zeta$, $r(w)\in\zeta$.  Two parallel sets $\xi$ and $\zeta$ are a {\em complete pair} if all paths $w$ that start in $\xi$  $s(w)\in\xi$ end up in $\zeta$, $r(w)\in\zeta$ and the reverse\footnote{In previous work, we used the terms {\em complete past} and {\em complete future} for complete source and complete range.}.   In that case, we write $\xi\leq \zeta$.  For such $\xi\leq\zeta$, 
 we have an
evolution of a closed quantum system and a unitary operator \beq
U\left(\xi,\zeta\right):{\cal H}(\xi)\rightarrow {\cal H}(\zeta).
\eeq This determines an isomorphism $\Phi(\xi, \zeta):{\cal
A}(\zeta)\rightarrow{\cal A}(\xi)$ via 
\beq
\Phi\left(\xi,\zeta\right)\left(\rho\right)=U\left(\xi,\zeta\right)\rho
\,
    U\left(\xi,\zeta\right)^\dagger\label{eq:qchU}\forall\rho\in{\cal A}(\xi).
\eeq Eq.\ (\ref{eq:qchphi}) is the restriction of (\ref{eq:qchU})
to ${\cal A}(x)\subseteq{\cal A}(\xi)$ for $x\in\xi$.  In more detail, we will see below that 
(\ref{eq:qchU}) can be reconstructed from the local maps
(\ref{eq:qchphi}) using the appropriate precise mathematical
definition of a quantum causal history\cite{HawMarSah}.

\begin{definition}
    A quantum causal history consists of a simple matrix \cs-algebra $\A(x)$ for every vertex $x\in V(\G)$ and a completely positive map $\m(x,y) : \A(y)\to\A(x)$ for every pair of related vertices $x\leq y$, satisfying the following axioms.
\end{definition}

\emph{Axiom 1:} (Extension) For any $y\in V(\G)$ and $\xi\subset
V(\G)$ a complete source of $y$, there exists a homomorphism\footnote{
Note that a completely positive map on density matrices is equivalent to a completely positive map of observable operators in the opposite direction.  A trace-preserving CP map of density matrices is dual to a unital CP map of observables, which is what we use in this definition.}
\[
\m_{\mathrm P}(\xi,y) : \A(y)\to\A(\xi)
\mbox,\]
such that for each $x\in\xi$, the reduction of $\m_{\mathrm P}(\xi,y)$ to
$\A(x)\subset\A(\xi)$ is $\m(x,y)$.
Likewise, for any $\zeta\subset \G$ a complete range of $y$, there
exists a map
\[
\m_{\mathrm F}(y,\zeta) : \A(\zeta)\to\A(y)
\mbox,\]
such that $\m^\dagger_{\mathrm F}(y,\zeta)$ is a homomorphism and for
any $z\in\zeta$, the reduction (restriction) of $\m_{\mathrm
F}(y,\zeta)$ to $\A(z)\to\A(y)$ is $\m(y,z)$. 

\emph{Axiom 2:} (Commutativity of unrelated vertices) If $y\sim z\in V(\G)$ and
$\xi\subset V(\G)$ is a complete source of $y$ and $z$, then the images of
$\m_{\mathrm P}(\xi,z)$ and $\m_{\mathrm P}(\xi,y)$ (in $\A(\xi)$)
commute.  Likewise, if $\zeta\subset \G$ is a complete target of
$\{x,y\}$, then the images of $\m_{\mathrm F}^\dagger(x,\zeta)$ and
$\m_{\mathrm F}^\dagger(y,\zeta)$ commute.

\emph{Axiom 3:} (Composition) If $\zeta$ is a complete source of $x$ and a complete range of $y$, then $\m(x,y) = \m_{\mathrm F}(x,\zeta)\circ\m_{\mathrm P}(\zeta,y)$.

If we are given the CP maps on the edges of the graph, we can recover the unitary operators between the complete pairs.  
This is because the axioms in the definition of a quantum causal history imply the following theorem \cite{HawMarSah}:
\begin{thm}\label{main}
    For any parallel sets  $\xi,\zeta\subset V(\G)$, if $\zeta$ is a complete
range of $\xi$ or $\xi$ is a complete source of $\zeta$ then there
    exists a unique map
    \[
    \m(\xi,\zeta):\A(\zeta)\to\A(\xi)
    \]
    such that
    \begin{enumerate}
	\item For any $x\in\xi$ and $z\in\zeta$, the reduction of
	$\m(\xi,\zeta)$ to $\A(z)\to\A(x)$ is $\m(x,z)$.
	\item If $\xi$ is a complete source of $\zeta$, then $\m(\xi,\zeta)$ is
	a homomorphism.
	\item If $\zeta$ is a complete range of $\xi$, then
	$\m^\dagger(\xi,\zeta)$ is a homomorphism.
	\item If $\xi\preceq\zeta$ is a complete pair, then $\m(\xi,\zeta)$ is
	an isomorphism.
\item If $\xi\preceq\upsilon\preceq\zeta$, then $\m(\xi,\upsilon)\circ \m(\upsilon,\zeta)=\m(\xi,\zeta)$.
    \end{enumerate}
\end{thm}

That is, the local information of the CP maps associated to edges implies the system of isomorphisms for complete pairs.
Note that, because $\G$ is assumed to be locally finite, there is no problem in going from  isomorphisms of algebras to a system of unitary maps; the only arbitrariness is a choice of irrelevant phase factors.

Also note that not every choice of unitary maps for complete pairs can be expressed in terms of CP maps in this way. This formulation enforces local causality when the edges of $\Gamma$ are interpreted as causal relations.

Definition 2 describes a simple structure, of open quantum systems that form a network.  It turns out that different insights can be extracted from it, depending on the interpretation of the graph $\Gamma$, mainly, and also the information contained in the $\Ai(x)$.  The rest of the paper describes the many lives of a quantum causal history:  as a discrete quantum field theory on a causal set (section \ref{sectionQFT}), a quantum geometry model for quantum gravity (section \ref{sectionQSF}) and, our main focus, a pre-spacetime quantum theory of gravity with no fundamental geometric degrees of freedom (section \ref{sectionQIP}).  

\section{QCH as a discrete quantum field theory}
\label{sectionQFT}

Local finiteness in the context of quantum field theory means a short distance cutoff on physical degrees of freedom.  This has raised a number of concerns.  As the universe expands, the number of degrees of freedom must be growing.  However, quantum field theory does not allow us to formulate dynamics when the number of degrees of freedom changes in time (see, for example, \cite{FosJac}  and references therein).  Semiclassical considerations of a deSitter universe, in the same locally finite context, suggest that we need a careful consideration of the quantum theory of that universe \cite{Banks}.  

To address such questions one may use a quantum causal history as a discrete version of algebraic quantum field theory. Algebraic quantum field theory (see \cite{Buc,Haa}) is a general approach to quantum field theory based on algebras of local observables, the relations among them, and their representations. The correct choice of axioms is still a matter of research, but for our
purposes, these subtleties are largely irrelevant.

Let $\Gamma$ be a partial order of events, the locally finite analogue of a Lorentzian spacetime.  Two events are causally related when $x\leq y$ and spacelike otherwise.  A parallel set $\xi$ becomes an {\em acausal set}, the discrete analogue of a spacelike slice or part of a spacelike slice.  The causal relation $\leq$ is transitive.  

An algebraic quantum field theory associates a von\,Neumann algebra to each causally complete region of spacetime. This generalizes easily to a directed graph.  
The following definitions are exactly the same as for continuous spacetime. For any subset $X\subset \G$, define the \emph{causal complement} as
\[
X' := \{y\in\G \mid \forall x\in X : x\sim y\}
\]
the set of events which are spacelike to all of $X$. The \emph{causal completion} of $X$ is $X''$, and $X$ is \emph{causally complete} if $X=X''$. A causal complement is always causally complete (i.e., $X'''=X'$).

In the most restrictive axiomatic formulation of algebraic quantum field theory there is a von\,Neumann algebra $\A(X)$ for every causally complete region.  These all share a common Hilbert space. Whenever $X\subseteq Y$, $\A(X)\subseteq\A(Y)$. For any causally complete region $X$, $\A(X')$ is $\A(X)'$, the commutant of $\A(X)$. The algebra associated to the causal completion of $X\cup Y$ is generated by $\A(X)$ and $\A(Y)$.

Some of the standard arguments about the properties of the local von\,Neumann algebras are valid for causal sets; some are not. The algebras should all be simple (i.e., von\,Neumann factors) because the theory would otherwise have local superselection sectors. For continuous spacetime it is believed that the local algebras should be type III${}_1$ hyperfinite factors; however, the reasoning involves the assumption that there exists a good ultraviolet scaling limit. This does not apply here; the small-scale structure of a causal set is discrete and not self similar at all.

Instead, the aim here (following the usual arguments for local finiteness) is that only a finite amount of structure should be entrusted to each event. In other words, each von\,Neumann algebra should be a finite-dimensional matrix algebra. In von\,Neumann algebra terms, these are finite type I factors.

Consider the causal completion $\xi''$ of a finite acausal set $\xi$;  $\A(\xi'')$ should be generated by the algebras $\A(x)$ for $x\in\xi$. In fact, these algebras should commute, and therefore
\[
\A(\xi'') =\A(\xi) = \bigotimes_{x\in\xi}\A(x)
.
\]
Evolution by homomorphisms is sometimes possible. If $\xi$ and $\zeta$ are acausal sets such that $\zeta\subset \xi''$, then $\zeta''\subseteq\xi''$ and so $\A(\zeta)\subseteq\A(\xi)$. This inclusion is the evolution homomorphism when $\xi$ is a complete past of $\zeta$.

Not surprisingly, simple matrix algebras are much easier to work with than type III von\,Neumann factors. Using the (unique) normalized trace, any state is given by a density matrix. Recall that the adjoint maps $\m^\dagger(x,y)$ in a quantum causal history are the induced maps on density matrices.

Suppose that $\rho\in\A(x)$ is a density matrix at $x$. The density matrix $\m^\dagger(x,y)(\rho)\in\A(y)$ is the best approximation to $\rho$ among density matrices at $y$, in the sense that it minimizes the trace norm of the difference. The trace norm metric on density matrices is equal to the metric on states induced by the operator norm on observables.

So, we see that the obvious notion of an algebraic quantum field theory on a causal set, with the physically reasonable assumption of finite algebras on events, gives the structure of a QCH. The two concepts intersect, but are not equivalent. 
One could also consider a field theory on $\Gamma$ with infinite algebras on events.  Depending upon the choice of algebras, even this might be described by some suitable generalization of a QCH. Likewise, a given QCH cannot necessarily be derived from an algebraic quantum field theory.

This means that the structure of a QCH encompasses a reasonable notion of a quantum field theory, and hence is capable of describing matter degrees of freedom. It also indicates how quantum fields on curved spacetime might be obtained as a limit of some quantum gravity model based on QCH's.

 This framework may be a good one to investigate questions like the ones in the beginning of this section.
 For example, CP evolution of finite-dimensional systems is widely used and well-understood in quantum information theory (see \cite{NieChu}), including the notion of probability and measurement in this context.  As for the puzzle of the growing number of degrees of freedom in an expansing universe, this is a good framework to study it, but we note that the puzzle assumes that the notion of growing is independent of these degrees of freedom.  If, instead, the fundamental theory of gravity is pre-geometric (as suggested in section \ref{sectionQIP}), this problem may be an artifact of the assumption that the local finiteness has been applied to the geometric degrees of freedom.


\section{QCH as a causal quantum geometry}\label{sectionQSF}

The construction in the previous section is of interest because it can be used to investigate some interesting questions that push the limits of applicability of quantum field theory but, just as with quantum field theory proper, it cannot be a quantum theory of gravity since the network $\Gamma$ of spacetime regions is not dynamical.  

A common path to a candidate quantum theory of gravity proposes that we need to consider a quantum superposition of geometries (as is the case in path integral quantum gravity, quantum Regge calculus and causal sets and more recently spin foams and Causal Dynamical Triangulations).  This can be done with a QCH in a straightforward way.  

Let  $\Gamma$ be a causal set, namely, a partial order of events that are causally related when $x\leq y$ and spacelike when $x\sim y$.  
A path integral quantum theory of gravity can be obtained from the superposition of all of the resulting quantum causal histories, that is, formally, the amplitude from the in state of the universe to the out state is
\beq
A_{{\mbox{\footnotesize{ in}}}\rightarrow {\mbox{\footnotesize{ out}}}}=\sum_\Gamma \prod_{e\in\Gamma}\Phi_e,
\label{eq:Ainout}
\eeq
where $\Gamma$ are all causal sets that match the boundary conditions of the in and out states\footnote{Due to lack of space, we will not discuss here the situation when the path-integral is the a projector from the kinematical to the physical states, as is the case in Loop Quantum Gravity.  The reader can consult the chapters by Perez and Oriti in this volume. }.

While, formally, eq.\ (\ref{eq:Ainout}) is the same as the causal set approach to quantum gravity, 
the introduction of the additional degrees of freedom at the level of the state spaces on the events on which the maps $\Phi_e$ act means that this is a {\em causal spin foam model}.   It appears that it is easier to extract information at the effective level when there is more input at this level, as we will see in section \ref{sectionCohDofsQGeo}.  In its basic steps, the construction of a causal spin foam as a QCH is as follows.  As this is only a sketch, for more details see \cite{MarWhe,Mar97,Ori}.  

To each event $x$, we associate an elementary Planck-scale state space of geometrical degrees of freedom.  
For example, this can be done as in {\em causal spin networks} \cite{MarSmo97,Mar97}.  Spin networks were originally
defined by Penrose as trivalent graphs with edges labelled by
representations of $SU(2)$ \cite{Pen}.  From such abstract labelled
graphs, Penrose was able to recover directions (angles) in
3-dimensional Euclidean space in the large spin limit.  Later, in Loop
Quantum Gravity, spin networks were shown to be the basis states for
the spatial geometry states.  The kinematical quantum area and volume operators,
in the spin network basis, have discrete spectra, and their
eigenvalues are functions of the labels on the spin network.  For a review of spin networks in LQG see \cite{LQG}.

Spin networks are graphs with directed edges labeled by representations of SU(2).  Reversing the direction of an edge means taking the conjugate representation.  A node in the graph represents the possible channels from the tensor product of the representations $\rho_{e_{\rm in}}$ on the incoming edges $e_{\rm in}$ to the tensor product of the representations on the outgoing ones, i.e., it is the linear map
\beq
\iota:\bigotimes_{e_{\rm in}} \rho_{e_{\rm in}}\rightarrow\bigotimes_{e_{\rm out}} \rho_{e_{\rm out}}.  
\eeq
Such a map $\iota$ is called an {\em intertwiner}.
The intertwiners on a node form a finite-dimensional vector space. 
Hence, a subgraph in the spin network containing one node $x$ 
 corresponds to a Hilbert space $H(x)$ of intertwiners.  
 That this is a finite-dimensional space, together with the local finiteness of $\Gamma$, ensures a locally finite theory.  
 Two 
spacelike events are two independent subgraphs, and the joint Hilbert 
space is $H(x\cup y)=H(x)\otimes H(y)$ if they have no common edges, 
or $H(x\cup y)=\sum_{\rho_{1},\ldots \rho_{n}}H(x)\otimes H(y)$, if $x$ and $y$  are joined in the spin network graph by 
$n$ edges carrying representations $ \rho_1,\ldots \rho_{n}$.    

Given an initial spin network, to be thought of as modeling a quantum
``spatial slice'', $\Gamma$ is built by repeated application of
local moves, local changes of the spin network graph.  Each move 
is a causal relation in the causal set.    The standard set of local generating moves for 4-valent spin networks is given by the following four operators:
\[
    \begin{array}{c}\mbox{\epsfig{file=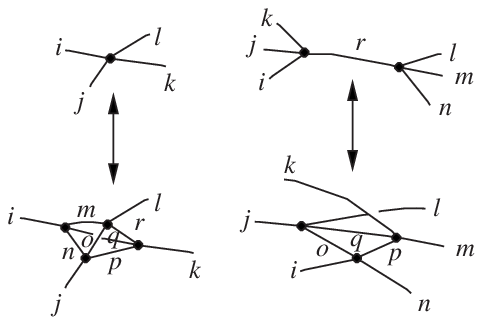}}\end{array}
\]
Note that the new subgraph has the same boundary as the original one and therefore corresponds to the same Hibert space of 
intertwiners.  A move is a unitary operator from a state 
$|S\rangle$ to a new one $|S'\rangle$ in $H$. 
Therefore, a causal spin network history is a causal set in which the 
vertices are the  Hilbert spaces of intertwiners and the causal relations are unitary 
operators\footnote{Or one may assign the state spaces to the edges of $\Gamma$ and the moves to the vertices.  Each of the two possible assignments has slightly different uses.  In any case, no preference either way can come from $\Gamma$ since for any given $\Gamma$ one can take its dual (mapping vertices to edges and edges to vertices) and hence the dual quantum causal history.}.   Inside these subgraphs the evolution is completely positive.  

A specific model of causal spin network evolution is given by a specific assignment of amplitudes to these moves, leading to an amplitude of the form 
\beq
A_{S_{\rm in}\rightarrow S_{\rm out}}=\sum_{\partial \Gamma = S_{\rm in}\rightarrow 
S_{\rm out}}\qquad
\prod_{{\mbox{\footnotesize moves}}\in\Gamma} A_{\mbox{\footnotesize{move}}}
\label{eq:Zcsn}
\eeq 
to go from initial spin network $S_{\rm in}$ to final spin network $S_{\rm out}$.

One should note that there is no preferred foliation in this model. 
The allowed moves change the network locally and any foliation
consistent with the causal set (i.e.\ that respects the order the
moves occured) is possible.  This is a discrete analogue of multifingered time evolution.  For more details, see \cite{Mar97}.

More sophisticated ways of obtaining a causal theory of quantum geometry are currently under development, using a causal version of group field theory (see Oriti, this volume).

\subsection{The low energy problem of background independent theories of quantum geometry}
\label{sectionLEP}

A microscopic model of spacetime is successful if it has a good low-energy limit in which it reproduces the known theories, namely general relativity with quantum matter coupled to it.  In the case of Causal Dynamical Triangulations \cite{CDT}, impressive results show strong indications that this model has several desired features.  This hinges on specific features of the model that allow a Wick rotation to a statistical sum and thus technical control of the sum via both analytic and numerical methods.  In the general spin foam case, obtaining any results on the low energy behaviour is a formidable problem.  In this section, we briefly discuss some basic aspects of the problem.  

Let us first note the similarity between the form of the model (\ref{eq:Zcsn}) and a condensed matter system.  Formally, the difference is that eq.\ (\ref{eq:Zcsn}) contains a quantum sum over all ``lattices''.  Physically, (\ref{eq:Zcsn}) is background independent while any condensed matter system has a scale since, of course, it lives in a fixed spacetime.  Although at first sight one may be intimidated by the first technical complication, it is really the second that is the killer.  
The low energy problem is indeed analogous to problems in condensed matter physics in that the aim is to describe  macroscopic  behaviour emergent from a many body system.  However, the  methods used in condensed matter physics, such as the renormalization group, rely on properties of the background geometry.  It is worth spending a few lines discussing exactly what issues one encounters.

The analogy to a condensed matter system should make us wary.  Eq.(\ref{eq:Zcsn}) is far more complicated than the kinds of systems that we can obtain results for in condensed matter physics.
Even if impossible  in practice, does this analogy bring us any conceptual insight?  
One of the most useful techniques used in condensed matter physics and quantum field theory to derive the macroscopic behavior of a system from the fundamental microscopic degrees of freedom is the renormalization group (RG).  
There has been quite some progress in applying RG ideas to spin foams in the past few years  and, at least in my opinion, the main result is that a background independent system is indeed very different than an ordinary system.  

To keep things simple, consider real space renormalization of a spin system, such as the coarse-graining of an Ising model.  Implicit in the method is the fact that, for a system in a fixed background, coarse-graining the lattice coarse-grains the observables.  In a background independent theory there is no direct relationship between BI observables and the lattice and, hence, coarse-graining the lattice does not guarantee coarse-grained observables.  

Furthermore, one has to deal with the superposition of lattices.  It is tempting to start by coarse-graining a single graph $\Gamma$ in the sum at a time but it is unclear what this could mean since it amounts to coarse-graining a single state. In the light of the previous comment, for this strategy to work, we would need to have a relationship between the observables and the graph of a single state.  Even if such a graph exists, finding it amounts to solving the low energy problem.  

Technically, it is possible to enlarge the definition of the RG to a sum over lattices.   In \cite{MarRG1} and \cite{MarRG2}, this was done by formulating the renormalization group in an algebraic setting in terms of Hopf algebras; this involved constructing a Wilsonian analogue of work of Connes and Kreimer \cite{ConKre}.  
The method works well on a fixed-background system on a sum-over-lattices but in a BI one we still face the problem of ordering couplings without knowing the observables.  

In a nutshell, the moral here is that a path integral of quantum geometries is a formal sum (with the exception of Causal Dynamical Triangulations) and there is little useful physics in a formal application of the renormalization group.  

Related to this is the fact that the low energy behaviour of any system is determined by its dynamics.  Dynamics is notoriously difficult to implement in any of the background independent approaches to quantum gravity.  In Loop Quantum Gravity, dynamics is the step from the kinematical to the physical sector.  The analogue of eq.(\ref{eq:Zcsn}) properly applied to LQG  is precisely this step.  Again, it is likely to be misguided  to apply any RG or similar methods to the kinematical states unless one has a reason to think that certain kinematical states already have the properties of the physical ones.   

In other approaches, such as causal sets, the usual problem is that one does not have explicit expressions for the  quantum dynamics and hence no physical couplings to analyze.  
It is possible that the recent progress in group field theory will lead to progress with the above issues (see the chapter by Freidel in this volume).  

It is very interesting to study how the low energy problem is dealt with in Causal Dynamical Triangulations (see the chapter by Loll).  Since the crucial step appears to be that CDT has a time direction, we may conclude that background independence can be handled as long as there is a well-defined causal direction.  Indeed, that is partly the ingredient  that we will use in the remaining of this article to address the low energy problem of background independent theories.  

Our proposal, developed in the next section,  is that, instead of looking for ways to coarse-grain the quantum geometry directly, one can first look for long-range propagating degrees of freedom and reconstruct the geometry from these (if they exist).  The specific method we adopt is promising because it deals directly with quantum systems and coarse-grains a quantum system to its effective particles.  However, we will find that, if geometry is indeed to be recovered from these effective degrees of freedom, there are no compelling reasons to start with a quantum sum over geometries for the microscopic theory (and its baggage of complications).  In fact, we will, in section  \ref{sectionQIP}, drop the microscopic geometric degrees of freedom and aim for a quantum theory of gravity in which gravity and geometry are only effective notions.  

But first, in the next section, we will present a definition of coherent effective degrees of freedom that can be used in an appropriately background independent theory.

\section{Noiseless Subsystems as effective coherent excitations}
\label{sectionNS}

Given the above considerations, we would now like to suggest an alternative path to the effective theory  of a background independent system.  
The basic strategy is to  begin by identifying {\em effective coherent degrees of freedom} and use these to  characterize the effective theory.  If these behave as if they are in a spacetime, we have a spacetime.  

In  \cite{MarPou,DreMarSmo,KriMar}, we suggested that to do this a suitable notion of
a coherent excitation from quantum information processing can be used.  This is the notion of a {\em noiseless subsystem} (NS) in quantum error correction, a subsystem protected from the noise, usually thanks to symmetries of the noise \cite{NS}.  Our observation is that passive error correction is analogous to problems concerned with the emergence and
stability of  persistent quantum states in condensed matter physics.     
In a quantum gravity context, the role of noise is simply the fundamental evolution and the existence of a noiseless subsystem 
means a coherent excitation protected from the microscopic
Planckian  evolution, and thus relevant for the effective theory\footnote{The term ``noiseless'' may be confusing in the present context:  it is not necessary that there is a noise in the usual sense of a given split into system and environment.  As is clear from the definition that follows, simply evolution of a dynamical system is all that is needed, the noiseless subsystem is what evolves coherently under that evolution.}.

\begin{definition}
Let $\Phi$ be a quantum channel on $\H$ 
and suppose that $\H$ decomposes as $\H =
(\H^A\otimes\H^B)\oplus\K$, where $A$ and $B$ are subsystems and
$\K =(\H^A\otimes\H^B)^\perp$. We say that $B$ is {\em noiseless}
for $\Phi$ if
\begin{eqnarray}\label{ns}
\forall\sigma^A\ \forall\sigma^B,\ \exists \tau^A\ :\
\Phi(\sigma^A\otimes\sigma^B) = \tau^A\otimes \sigma^B.
\end{eqnarray}
\label{eq:NS}
\end{definition}
Here we have written $\sigma^A$ (resp.\ $\sigma^B$) for operators
on $\H^A$ (resp.\ $\H^B$), and we regard $\sigma = \sigma^A\otimes
\sigma^B$ as an operator that acts on $\H$ by defining it to be
zero on $\K$.  Note that, given $\Hi$ and $\Phi$, it is a non-trivial problem to find a decomposition that exhibits a NS.  Much of the relevant literature in quantum information theory is concerned with algorithmic searches for a NS given $\Hi$ and $\Phi$.  

The noiseless subsystem method (also called decoherence-free
subspaces and subsystems) is the fundamental passive technique for
error correction in quantum computing.  In this setting, the operators $\Phi = \{E_a\}$ in the
operator-sum representation for a channel are
called the {error} or {noise} operators associated with
$\Phi$. It is precisely the effects of such operators that must be
mitigated for in the context of quantum error correction \cite{NS}.
The basic idea in this setting is to (when
possible) encode initial states in sectors that will remain immune
to the deleterious effects of the errors $\Phi = \{E_a\}$
associated with a given channel.

For our purposes, when $\Phi$ and $\Hi$ are a quantum causal history or other candidate quantum theory of gravity, the NS is a subsystem emergent (protected) from the microscopic
Planckian  evolution.

It will be useful to give the following necessary and sufficient condition for the existence of a noiseless subsystem. 
Consider the same  quantum system of state space $\Hi$ undergoing evolution by a completely positive operator $\Phi$.  
From the discussion of \ref{sectionCP}, the operator-sum representation of $\Phi$ acting on a density matrix $\rho\in\Hi$ is
\beq
\Phi\left[\rho\right]=\sum_a E_a\rho E_a^\dagger.
\eeq
If ${\cal B}\left({\cal H}\right)$ is the algebra of all operators acting on ${\cal H}$, the {\it evolution algebra} $\Aevol\subseteq {\cal B}\left({\cal H}\right)$ is the subalgebra generated by the $E_a$ (assuming that $\Aevol$ is closed under $\dagger$ and $\Phi$ is unital).  Up to a unitary transformation, $\Aevol$ can be written as a direct sum of $d_j\times d_j$ complex matrix algebras, each of which appear with multiplicity $\mu_j$:
\beq
\Aevol\simeq  \bigoplus_j {\one}_{\mu_j}\otimes{\cal B}\left({\bf C}^{d_j}\right),
\eeq
where ${\one}_{\mu_j}$ is the identity operator on ${\bf C}^{\mu_j}$.

The {\it commutant}, $\Aevol'$ of $\Aevol$ is the set of all operators in ${\cal B}\left({\cal H}\right)$ that commute with every element of $\Aevol$, i.e.,
\beq
\Aevol'\simeq\bigoplus_j {\cal B}\left({\bf  C}^{\mu_j}\right)\otimes{\one}_{d_j}.
\label{eq:Adecomp}
\eeq
This decomposition induces a natural decomposition of ${\cal H}$:
\beq
{\cal H}=\bigoplus_j{\bf C}^{\mu_j}\otimes{\bf C}^{d_j}.
\label{eq:c}
\eeq
Note now that any state $\rho$ in $\Aevol'$ is a fixed point of $\Phi$ since it commutes with all the $A_k$:
\beq
\Phi\left[\rho\right]=\sum_a E_a\rho E_a^\dagger=\sum_a E_a E_a^\dagger\rho=\rho.
\eeq
It can be shown that the reverse also holds, i.e., 
\beq
\Phi\left[\rho\right]=\rho\quad\Leftrightarrow\quad\rho\in\Aevol'.
\eeq
Hence, the noiseless subsystem can be identified with the ${\bf C}^{\mu_j}$ in equation (\ref{eq:c}).  
In the rest of this article we will use both the definition (\ref{eq:NS}) and the commutant to characterize the effective coherent degrees of freedom, depending on which one is easier to apply.  

A non-trivial commutant means non-trivial noiseless subsystems (i.e.\ symmetries in $\Aevol$ that result in protected degrees of freedom).  An important aspect of this method for quantum gravity is that, since the protected degrees of freedom are in the commutant of the evolution algebra,   it suffices to know what {\em kind} of operators can act on the system during evolution even when we do not know the precise dynamics. 

In what follows, we will investigate the idea that ``we have a spacetime if the coherent excitations
behave as if they are in a
spacetime''  in detail.  In section \ref{sectionSymmetry}, we will  think of the NS as particles even though, at this level, there is no spacetime and thus the usual notion of particles (as in Wigner) does not apply.  
Then, to have a Minkowski spacetime, we need the coherent excitations and 
 their interactions to be invariant under Poincar\'{e}
transformations. This
turns around the usual order:  a particle is not Poincar\'{e}
invariant because it is in a Minkowski spacetime, rather, all we
can mean by a Minkowski spacetime is that all coherent degrees of
freedom and their interactions  are Poincar\'{e} invariant at the
relevant scale.

\subsection{Coherent excitations in a theory of local moves on a graph}
\label{sectionCohDofsQGeo}
For this example, we consider a theory based on graph states evolving under local moves, as in Section \ref{sectionQSF}.  The state space of such a theory has the form
\beq
\Hi=\bigoplus_S\Hi_S,
\eeq
where the direct sum runs over all allowed spin networks $S$ and the state space $\Hi_S$ for each spin network is 
\beq
\Hi_S=\sum_{\{\rho\}}\bigotimes_{n\in S}\Hi_n. 
\label{eq:Hold}
\eeq
$n$ are the nodes of  $S$ and $\Hi_n$ is the intertwiner space for the subgraph of $S$ that contains node $n$.  The sum is over the representations on the edges connecting the nodes\footnote{Note that it is  not important that the state space is based on spin networks.  Any graph theory with a state space of the form \ref{eq:Hold} and evolution moves that are local in this decomposition will do.}. 

The move dynamics are generated by four operators corresponding to the four moves 
 illustrated in section \ref{sectionQSF}, which we will name $A_i, i=1,..., 4$, acting via 
\beq
A_i|S\rangle=\sum_\alpha|S_{\alpha i}'\rangle.
\eeq
$S_{\alpha i}'$ are the graphs obtained from $S$ by an application of one move of type $i$. 
Together with the identity, these moves generate the evolution algebra 
\beq
\Aevol=\left\{ {\one}, A_i\right\}
\eeq
on $\Hi$.

Our basic idea is that, if there is an effective theory, it is characterized by effective degrees of freedom which remain largely coherent, i.e., protected from the Planckian evolution and hence relevant for the low energy limit.  It is easiest to search for such excitations in an idealized setup in which they are left completely invariant under the evolution generated by $\Aevol$.  

Are there any such non-trivial excitations in $\Hi$?   There are, and are revealed when we rewrite $\Hi_S$ in eq.\ (\ref{eq:Hold}) as\footnote{In general, finding the decomposition of the system that reveals the noiseless subsystem is a non-trivial task and a large part of the literature on passive quantum error correction is devoted to finding algorithms that produce the desired decomposition. } 
\beq
\Hi_S=\Hi_S^{n'}\otimes\Hi_S^B,
\label{eq:Hnew}
\eeq
where $\Hi_S^{n'}:=\bigotimes_{n'\in S}\Hi^{n'}$ contains all {\em unbraided} single node subgraphs in $S$ (the prime on $n$ serves to denote unbraided) and $\Hi_S^b:=\bigotimes_{b\in S}\Hi_b$ are state spaces associated to braidings of the edges connecting the nodes.  For the present purposes, we do not need  to be explicit about the different kinds of braids that appear in $\Hi_S^b$.  

The difference between the decomposition (\ref{eq:Hold}) and the new one (\ref{eq:Hnew}) is best illustrated with an example.  Given the state
\[
    \begin{array}{c}\mbox{\epsfig{file=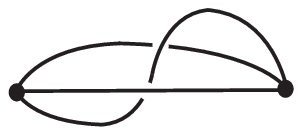}}\end{array}
\]
eq.\ (\ref{eq:Hold}) decomposes it as 
\[
    \sum_{\{\rho\}}
    \begin{array}{c}\mbox{\epsfig{file=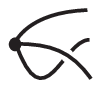}}\end{array}\otimes
     \begin{array}{c}\mbox{\epsfig{file=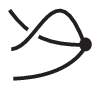}}\end{array}
\]
while (\ref{eq:Hnew}) decomposes it to 
\[
    \sum_{\{\rho\}}
    \begin{array}{c}\mbox{\epsfig{file=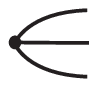}}\end{array}\otimes
    \begin{array}{c}\mbox{\epsfig{file=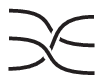}}\end{array}\otimes
     \begin{array}{c}\mbox{\epsfig{file=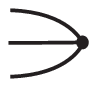}}\end{array}.
\]

With the new decomposition, one can check that operators in $\Aevol$ can only affect the $\Hi_S^{n'}$ and that $\Hi_S^b$ is {\em noiseless under} $\Aevol$.  This can be checked explicitly by showing that the actions of braiding of the edges of the graph and the evolution moves commute.  

We have shown that braidings of graph edges are unaffected by the usual evolution moves.  Any physical information contained in the braids will propagate coherently under $\Aevol$.  These are effective coherent degrees of freedom.  

One may worry that braids are topological degrees of freedom.  That is, the middle subgraph in the above tensor product decomposition appears localized in this particular embedding but a diffeomorphism of the graph could spread the braid over the entire graph.  The braid remains a coherent degree of freedom but it is not localized with respect to $\Aevol$ (i.e., the topology defined by the graph).  This could lead us to conclude that such effective degrees of freedom are not the ones we are looking for.  However, more careful analysis shows that this conclusion is naive and leads to interesting new directions, as we will see in section \ref{sectionlocality}.


\section{QCH as pre-spacetime quantum information processor}
\label{sectionQIP}

In section \ref{sectionNS}, we argued that the effective theory of a physical system, including a background independent one,  can be characterized by the behaviour of the effective degrees of freedom.  If, then,  general relativity is an effective theory, it should also be characterized by effective degrees of freedom.
Since non-perturbative gravity is equivalent to dynamical spacetime geometry, what these effective degrees of freedom need to do is to characterize geometry and hence explain gravity.  That is, geometry needs to appear at the effective level. 
From this perspective, it is unnecessary and circular to have fundamental degrees of freedom that are geometric.  

In this section, we begin to formulate a new approach to quantum gravity, one that is both background independent and has no fundamental gravitational or geometric degrees of freedom, 
a setup that has not been investigated except for recent work in \cite{Dre,Llo}.  We will see that our favorite framework, QCH, can mutate once more to help us define the problem.  

Let us first state the definition of background independence we will use\footnote{
A longer discussion of background independence in this context can be found in \cite{KriMar}.  Background independence is discussed in great detail in \cite{ButIsh,Sta,Smo}.
}: 

\begin{definition}
A quantum  theory of gravity is {\em background independent} if its basic quantities and concepts do not presuppose the existence of a given background metric.
\end{definition}

The key idea now is that 
unnecessary references to a background spacetime can be eliminated by using the language of quantum information processing. 
The new use of a QCH is to read it as a background independent quantum information processing system and derive directly  the  emergent degrees of freedom.  Interactions of those are the events of our spacetime.  There are no separate gravitational degrees of freedom to be quantized.   

The algebras $\Ai(x)$ on the vertices $x$ of $\Gamma$ now are quantum systems with no spatiotemporal attributes and $\Gamma$ is the circuit of information flow between them (the collection of relevant completely positive maps).  Such circuits, as used in quantum information theory,  are directed graphs with no cycles, precisely the definition of $\Gamma$ in section \ref{sectionGamma}.   In this form, the QCH is suitable for the application of the method of section \ref{sectionNS}  that introduces geometrical properties at the level of effective coherent degrees of freedom encoded in the system.   It is important to note that the effective degrees of freedom will not have a causal structure directly related to $\Gamma$ (think of the braid example in section \ref{sectionCohDofsQGeo} where the braids do not interact at all and form a trivial causal structure no matter how complicated the flow of graph moves may be).  

The aim is to find features of a classical geometry, such as symmetries, at the level of these effective degrees of freedom, without starting with a quantum geometry.  In the next section, we take a first step towards deriving geometry from such a setup.  We suggest that the coherent excitations  ``behave as if they are in a
spacetime'' if they and their interactions are invariant under Poincar\'{e}
transformations. 


\subsection{Spacetime via the symmetries of the excitations}
\label{sectionSymmetry}

In this section, we provide a possible suggestion of how aspects of a spacetime can emerge from the excitations (as in section \ref{sectionCohDofsQGeo}) of a pre-spacetime quantum system (as in section \ref{sectionQIP}), by a rather straightforward and literal interpretation of the idea that all we
can mean by a Minkowski spacetime is that all coherent degrees of
freedom and their interactions  are Poincar\'{e} invariant at the
relevant scale.  

We follow \cite{KriMar}, where this is implemented in an idealized setup where no relevant scale is introduced, instead, the coherent degrees of freedom are invariant at all scales.  This is done by a generalization of the definition of a noiseless subsystem to a {\em group-invariant noiseless subsystem}, as follows. 

Let $G$ be a group and $\pi:G\rightarrow\B(\H^{\rm
rep})$ a (unitary) representation of $G$ on 
$\H^{\rm rep}$. We identify $G$ with the unitary group $\pi(G)$.
For each $U$ in $G$ denote the corresponding superoperator on
$\B(\H)$ by $\U(\cdot) = U(\cdot)U^\dagger$. To simplify
the notation below, we denote the representation Hilbert
space as $\H^B\equiv\H^{\rm rep}$. As previously, we are interested in scenarios
for which $\H^B$ is a subsystem of a larger Hilbert space $\H$, that is, $\H$ decomposes as $\H =
(\H^A\otimes\H^B)\oplus\K$, where $\K = (\H^A\otimes\H^B)^\perp$.
Again, $\Phi$ is a quantum channel on $\H$.

\begin{definition}
{ We say that $B$ is {\em group invariant under $\Phi$} if there
is a $U$ in $G$ such that
\begin{eqnarray}\label{maineqn}
\forall\sigma^A\ \forall\sigma^B,\ \exists \tau^A\ :\
\Phi(\sigma^A\otimes\sigma^B) = \tau^A\otimes \U(\sigma^B).
\end{eqnarray}
}
\end{definition}
The groups of interest are,  of course, Poincar\'{e}, deSitter, etc\footnote{ Similar constructions, in a different context, can be found  in the quantum information literature.  See \cite{KriMar} for details.}.

The following theorem gives a number of testable conditions that
are equivalent to group invariance of a noiseless subsystem. Namely, the result shows how
this notion may be  phrased in terms of the partial trace
operation on $A$; that it is enough to satisfy this equation for
the maximally mixed state on $A$; of practical interest, condition 4 shows {\em how to test if a given subsystem
satisfies this equation if a choice of operator elements for the
evolution map is known}.  The last condition is the corresponding statement in terms
of operator algebras.   

Let $P^{AB}$ be the projection of $\H$ onto
$\H^A\otimes\H^B$ and define a ``compression superoperator''
$\P^{AB}(\cdot) = P^{AB}(\cdot)P^{AB}$ on $\H$. That is, $\P^{AB}$
is the map on $\B(\H)$ defined by $\P^{AB}(\sigma) = P^{AB}\sigma
P^{AB}$, $\forall\sigma\in\B(\H)$. Then in terms of the partial
trace operation on $A$, Eq.~(\ref{ns}) is equivalent to the
statement
\begin{eqnarray}\label{ns1}
\Tr_A \circ \Phi \circ \P^{AB} = \Tr_A \circ \P^{AB}.
\end{eqnarray}
With this definition, the theorem in \cite{KriMar} states:

\begin{thm}\label{thm:NS}
Let $G$ be a group represented on a Hilbert space $\H^B$. Suppose
that $\H$ is a Hilbert space that decomposes as $\H =
(\H^A\otimes\H^B) \oplus\K$, and that
$\Phi:\B(\H)\rightarrow\B(\H)$ is a quantum channel. Then the
following five conditions are equivalent:
\begin{itemize}
\item[1.] $B$ is  group invariant under $\Phi$. \item[2.] $\exists\,
U\in G : \forall\sigma^B,\ \exists \tau^A\ :\
\Phi(\one^A\otimes\sigma^B) = \tau^A \otimes \U( \sigma^B)$
\item[3.] $\exists\, U\in G :\ \Tr_A\circ \P^{AB}\circ \Phi\circ
\P^{AB} =\U\circ\Tr_A\circ\P^{AB}$. \item[4.] Let
$\{\ket{\alpha_k}\}$ be an orthonormal basis for $\H^A$ and let
$\{P_{kl} = \kb{\alpha_k}{\alpha_l}\otimes\one^B \}$ be the
corresponding family of matrix units in $\B(\H^A)\otimes\one^B$.
Let $\Phi=\{E_a\}$ be a choice of operator elements for $\Phi$.
Then there is a $U\in G$ such that
\begin{equation}
P_{kk} (\one^A\otimes U^\dagger) E_a P_{ll} = \lambda_{akl} P_{kl}
\quad\forall\, a,k,l \label{eq:cond1}
\end{equation}
for some set of scalars $\{\lambda_{akl}\}$ and
\begin{equation}
 (\one^A \otimes U^\dagger) E_a P^{AB} = P^{AB}  (\one^A \otimes U^\dagger)
 E_a P^{AB} \quad\forall\, a. \label{eq:cond2}
\end{equation}
\item[5.] There is a $U\in G$ such that the subspace
$\H^A\otimes\H^B$ is invariant for the operators $\{ (\one^A
\otimes U^\dagger) E_a \}$, and the restricted operators $\{
(\one^A \otimes U^\dagger) E_a P^{AB}\}$ belong to the operator
algebra $\B(\H^A)\otimes\one^B$.
\end{itemize}
\end{thm}

Work is currently in progress towards the reverse of condition 4, namely, an algorithmic procedure that constructs the class of dynamics that contains the desired subsystems.  This is of interest because it amounts to  reverse-engineering a class of microscopic theories containing interesting geometric coherent degrees of freedom.  
Work is also in progress that incorporates an appropriate notion of group invariant interactions to the interactions of the group invariant noiseless subsystems.

Let us, in this context, note some of the most interesting features of noiseless
subsystems in a quantum gravity context.  First, they are not
localized, thus their symmetry is global. This is central to the discussion of microscopic
versus emergent locality in quantum gravity that follows in the next section. They illustrate the
fact that the emergent degrees of freedom can bear little relation
in their interactions to the underlying microscopic theory, known
of course from condensed matter physics, but now in a manifestly
background independent form. Second, the construction employs
quantum channels, rather than a partition function of the usual
spin foam type, which applies both to a single underlying circuit
(or history) or to a path integral sum\footnote{It is also of interest that our results can be applied to spin
foams with a boundary to extract the particles they contain and
thus address the outstanding low energy issue of these models.
The boundary is needed because the current definition of a noiseless subsystem requires a true time evolution (as opposed to a constraint).  This, and ways around it, are discussed in \ref{sectionTime}.
The importance of the boundary is also emphasized in
\cite{Carlo}.}.  Finally, it is very
important that the existence and properties of the noiseless
subsystems depends entirely on the properties of the dynamics.  As
can be seen in the quantum information literature and in their application to quantum gravity  in
concrete examples of noiseless subsystems their existence depends
on having symmetries in the dynamics.  We do know from standard physics that the effective description of a system depends on its dynamics, but it has been difficult to implement this in a background independent system.  

The following are shortcomings in the current application  of noiseless subsystems to quantum gravity:   The group invariant noiseless subsystems are not truly emergent but encoded in the microscopic dynamics, in the sense that both the symmetries of the dynamics that guarantee the existence of the noiseless subsystems and the group are present in the microscopic dynamics.  One would like to extend the relevant notions to an appropriate definition of an approximate, emergent group invariant noiseless subsystem.  
Also, there is no role for gravity here.

First, in the following section, we will discuss issues related to the notion of locality when spacetime is to emerge from a background independent fundamental theory.   In this case, there are two notion of locality, the emergent one in the emergent spacetime and that of the fundamental theory.  It is important to note that the two will likely not agree, as discussed next.

\subsection{Emergent locality in background independent quantum gravity}
\label{sectionlocality}

In this section, we will see that the perspective of obtaining an effective spacetime via the excitations of a pre-spacetime background independent theory allows us to make a preliminary analysis of locality issues in background independent quantum gravity.  Some of the issues encountered here have already appeared in other approaches to quantum gravity but as problems with no obvious resolution.  Here we will argue for a new kind of quantum gravity phenomenology that may be radical in its basis but, if this general direction is correct, the resulting quantum gravity effects will not be restricted to Planck scale corrections.

It should come as no surprise that locality is a tricky issue in BI quantum theories of gravity.  There is no background metric with which to measure distances or intervals.  And, as is well-known,  it is non-trivial to construct diffeomorphism invariant observables that measure local properties of fields. 
This is what we already know from general relativity, our classical background independent theory.  

For quantum gravity, the expectation that spacetime is to emerge from a background independent fundamental theory means that 
there are two notions of locality that may be relevant.  In a simple generic setup in which the underlying theory is given by some network of quantum systems as in section \ref{sectionQIP} one finds the following.  
In a given graph (the fundamental theory)  there will be a notion of locality:  in a graph two nodes are neighbors if they are connected by a link.
We can call this {\it microlocality.}  In the known background independent theories, the dynamics is generated by moves that are local in this microscopic sense.  
But if this is to be a good theory, there should be a notion of classical spacetime geometry that emerges from the quantum geometry.  This will give rise to another notion of locality,  which we may characterize as {\it macrolocality.}

The question is then whether there is any guarantee that these two notions of locality will coincide. This seems indeed unlikely.  In a theory if the type of section \ref{sectionQIP}, we see from the known examples of effective coherent excitations that they do not coincide.  Take the braid example:  the braids are global in the network and, since they are exactly conserved and do not interact, the macrolocality defined by their trivial dynamics is always the same trivial one independently of the microlocality.  In a very different example, Loop Quantum Gravity, it is again unlikely that micro and macrolocality will coincide, given that  the quantum states which are expected to represent classical spacetimes are to be constructed from superpositions of graph states each of which carries its own notion of locality. 

The noiseless subsystem viewpoint we have used here  brings in new physical understanding of this question. 
There is no reason that the microlocality coincides with the macrolocality of the effective spacetime.  Instead, the notion of macrolocality should be  defined directly from the interactions of the noiseless subsystems that we identify with the emergent degrees of freedom (elementary particles).  
Then a new possibility appears of a dual viewpoint:  the locality of the spacetime is to be given and identified by the effective degrees of freedom.  {\em It is the fundamental evolution that is non-local with respect to our spacetime. }

 We may note that this scenario is very different from the one that has been commonly assumed in many discussions about how space and time are to emerge from background independent models of quantum geometry. It has been a common assumption that
the quantum geometry describes  a ``bumpy'' classical geometry so that the 
microlocality of the evolution ought to coincide with that of the effective spacetime up to Planck scale corrections.

The test of this approach will be whether it can be worked out in detail.  
There are very promising candidate models for a phenomenology of this sort (such as \cite{Wen}) as well as promising examples of large scale non-locality (such as the CMB).   

For completeness, we note that the causal set approach also finds (with rather different arguments) that a fundamental non-locality may be desirable for a good low energy limit, Causal Dynamical Triangulations have a gauge condition which can be seen as a global constraint in each slice \cite{MarSmoCDT}, and there is a very rough similarity between this discussion and  ideas in brane-worlds and the AdS/CFT correspondence.  

\subsection{Time}\label{sectionTime}

The problem of time in quantum gravity appears again in the low energy problem, as we discussed in section \ref{sectionLEP}.  Ultimately, the low energy problem needs a resolution of the problem of observables in a background independent theory and observables are hard to come by in a constrained theory. 

The idea that the effective spacetime is to be defined from coherent excitations of a background independent quantum system suggests a new way to address this old problem:  it may be the constrained system is extracted from an underlying theory with time. In the light of the discussion in the previous section, we can see that the time of the underlying theory is likely not the same as the time of the effective one:  the above discussion on the discrepancy between underlying and effective locality of course goes through also for underlying vs effective causality.  Hence, it may be possible to have an underlying (``micro'') time without running into the observationally excluded preferred frame.  

In \cite{KonMar}, this idea was investigated in the context of the passive quantum correction method to extract the effective coherent excitations.  
One can understand the origin of physical states 
that solve all constraints as those states spanning noiseless 
subsystems in the Hilbert space of a particular system, when that 
system is coupled to an external environment. 

Consider a system $\Hi_S$ coupled to a bath $\Hi_B$ so that the total state space is $\Hi_{full}=\Hi_S\otimes\Hi_B$.  The evolution of the system is given by a hamiltonian 
\beq
H_{full} = H_S\otimes  {\one}_B + 
 {\one}_S\otimes H_B + H_I,
\eeq
where the interaction term $H_I$ can be decomposed as 
\beq
H_I=\sum_\alpha N_a\otimes B_a,
\label{Hint}
\eeq
 with  $N_\alpha$ and $B_\alpha$ operators 
acting on the system and bath respectively. $H_S, {\one}_S$ and the $N_\alpha$ generate an algebra $\Ai$ that decomposes as in eq.\ (\ref{eq:Adecomp}) with the resulting decomposition (as in eq.\ (\ref{eq:c})) of $\Hi_S$ onto noiseless and noisy sectors: 
\[
{\cal H}=\bigoplus_j{\bf C}^{\mu_j}\otimes{\bf C}^{d_j}.
\]
An interesting specialization of noiseless subsystems for operators $N_\alpha\in\Ai$, acting on $|a,b\rangle$, where $a$ and $b$ denote states according to the above decomposition,
is when
\beq
N_\alpha|a,b\rangle=p_{ab}|ab\rangle, 
\eeq
with the $p_{ab}$ being simply phases.    In such special cases, the operators $N_\alpha$ are called  {\em stabilizers}.

Next, consider a system subject to a set of first-class constraints 
$C_a$, acting on the kinematical state space $\Hi_{kin}$. 
The physical state space $\Hi_{phys}$ contains the states $|\psi_{phys}\rangle$ such that $C_\alpha|\psi_{phys}\rangle=0$. 

If $ {\one}$ is the identity operator on  ${\Hi}_{kin}$, define new operators $N_{a\, 
\lambda}=( {\one}+\lambda C_a)$. Then if $C_a|\psi\rangle_{phys} = 0$, the
operators $N_{a\, \lambda}$ stabilize physical states for all 
$\lambda$, $N_{a\, \lambda}|\psi\rangle_{phys} = 
|\psi\rangle_{phys}$. Thus, an alternative description of the 
constrained system starts to develop in which ${\Hi}_{kin}$ can 
be identified with $\Hi_{S}$ and the new stabilizer elements 
$N_{a\, \lambda}$ generate the algebra ${\Ai}$. Recall that 
elements of ${\Ai}$ have the interpretation of being operations 
that that couple the system to an environment. Thus, this 
approach suggests ${\Hi}_{kin}$ should be coupled to a new 
Hilbert space ${\Hi}_B$ representing an environment. 

The interaction Hamiltonian for the constrained 
system and environment now takes the form
 \beq H_I = \sum_{a} N_a 
\otimes B_a = \sum_a \left( 1 \otimes B_a + \lambda C_a \otimes 
B_a \right),\eeq for some operators $B_a$ acting on the 
environment.
 Only the terms proportional to 
the constraints are therefore part of the `true' interaction 
Hamiltonian, \beq \label{newHint} H_I \rightarrow \sum_a C_a \otimes B_a. 
\eeq In short, what we now have is a new quantum system with a 
full Hilbert space $\Hi_{full}=\Hi_{kin}\otimes \Hi_B$ 
governed by a Hamiltonian of the form (\ref{Hint}) with $H_S$ 
given by the Hamiltonian of the constrained problem, $H_B$ given 
by the operators $B_a$, and $H_I$ given by (\ref{newHint}). 

The noiseless states of this new theory are, by construction, 
solutions to the constraints $C_a$ that we started with. They, 
therefore, exhibit all the physical properties that the solutions to 
the constrained problem do. Since the environment in the quantum 
information theoretic description is an extra tool and is not 
really of interest from the point of view of the constrained 
dynamics problem, it should be traced out. As a result, the 
noiseless states evolve unitarily under the full Hamiltonian 
while the noisy states, which do not satisfy the constraint 
equations, decay non-unitarily and as such are not of physical 
interest.\footnote{The commutant $\Ai^\prime$ in the noiseless subspace picture 
is the set of all operators that commute with the constraints 
$C_a$. Thus, there is also a close correspondence between 
$\Ai^\prime$ and the Dirac algebra, up to the status of the unit 
operator. The unit is technically always a Dirac observable and 
is thus in the Dirac algebra. On the noiseless subsystem side, however, 
the unit operator is included in the algebra $\Ai$ (recall 
that $\Ai$ is assumed unital). Therefore, strictly speaking, 
the correspondence is between $\Ai^\prime$ and the set of 
non-trivial Dirac observables.}

Note that noiseless states evolve as if the Hamiltonian were zero 
exactly as in the original constrained system. Thus, these states 
exhibit an emergent time-reparametrization invariance property. 
(For example, the relativistic particle  is to be viewed in a similar manner as 
an excitation over a noisy background.)
However, in the noiseless subsystems picture, the `true' 
Hamiltonian is actually $H_{full}$ and is nonzero. There is no 
`problem of time' as the evolution of the environment provides a 
well defined mechanism to measure time flow by. 

This viewpoint is orthogonal to the much 
discussed relational approach where the introduction of a background 
time is seen as something that should be avoided. The relational idea is that time can arise from a timeless relational theory.  In contrast, \cite{KonMar} argues in the reverse 
direction that the relational features usually ascribed to 
physical systems such as the relativistic particle can be 
understood as arising out of a non-relational theory of the 
system under consideration and its environment. 

Here,
symmetries such as gauge invariance, diffeomorphism invariance, 
and time re-parametrization invariance are not fundamental 
features of the full system comprising the various environments. 
Adding an environment to the universe is certainly a strange move 
with interpretational issues if the quantum theory of gravity is simply the quantization 
of the known gravity and matter.  In that case the noisy states are unphysical.  
However, the situation is different in quantum gravity approaches, 
such as in section \ref{sectionQIP} or the
condensed matter approaches discussed in the chapter by Dreyer in this volume, 
in which general relativity is expected to be an effective theory of the excitations of a system with no fundamental gravitational degrees of freedom.
It should also be noted that coupling gravitational degrees of freedom to an external environment is, in fact, common; the Hilbert space of an FRW spacetime, for example, can be coupled to a scalar field to model inflation.

An answer to the problem of time may then be 
 that general relativity is the noiseless sector of the underlying quantum theory of gravity.

\subsection{Gravity for free?}
\label{sectionConjecture}

If spacetime is emergent, what is the role of the Einstein equations?  The new direction described here leads to a new approach on this problem because we use 
 the excitations and their interactions to define {\em both} the geometry and the energy-momentum tensor $T_{\mu\nu}$.   This leads to the following Conjecture on the role of General Relativity (formulated in collaboration with O.\ Dreyer, see also Dreyer, this volume): 

 {\em If the assignment of geometry and $T_{\mu\nu}$ from the same excitations and interactions is done consistently, the geometry and $T_{\mu\nu}$ will not be independent but will satisfy Einstein's equations as identities. }
 
The long-term goal of this direction would then be to realize this conjecture in the context of specific models of quantum spacetime.

\section{Conclusions}

In this article we started with the traditional background independent approaches to quantum gravity which are based on quantum geometric/gravitational degrees of freedom.   We saw that, except for the case of causal dynamical triangulations, these encounter significant difficulties in their main aim, i.e., deriving general relativity as their low energy limit.  We then suggested that general relativity should be viewed as a strictly effective theory coming from a fundamental theory with no geometric degrees of freedom (and hence background independent in the most direct sense).  

The basic idea is that an effective theory is characterized by effective coherent degrees of freedom and their interactions. Having formulated the pre-geometric BI theory as a quantum information theoretic processor, we were able to use the method of noiseless subsystems to extract such coherent (protected) excitations.  We followed the consequences:  truly effective spacetime means effective locality and effective time direction that are not simply Planck scale quantum corrections on the classical ones.  In particular, the discrepancy between the effective and the fundamental locality suggest a new direction in quantum gravity phenomenology.  

 I believe this  is very promising for three reasons:  1) The emphasis on the effective coherent degrees of freedom addresses directly and in fact uses the dynamics.  The dynamics is physically essential but almost impossible to deal with in other approaches.  2) A truly effective spacetime has novel phenomenological implications not tied to the Planck scale which can be tested and rejected if wrong.  3) A pre-spacetime background independent quantum theory of gravity takes us away from the concept of a quantum superposition of spacetimes which can be easily written down formally but has been impossible to make sense of physically in any approach other than Causal Dynamical Triangulations.  

Some of the more exciting possibilities we speculated on included solving the problem of time and deriving the Einstein equations.  Clearly this direction is in its beginning, but the basic message is that taking the idea that general relativity is an effective theory seriously involves rethinking physics without spacetime and is likely to have large scale consequences.  
This opens up a whole new set of possibilities and opportunities.  

\section*{Acknowledgments}
I am grateful to Olaf Dreyer, Tomasz Konopka, David Kribs, David Poulin and Lee Smolin for extensive collaborations and discussions during which several of the ideas presented here were developed.  
At various stages of this work, I benefited from discussions with and suggestions from  Robert Brandenberger, Steve Carlip, Ray Laflamme, Seth Lloyd, John Stachel, Paolo Zanardi and many of my colleagues at the Perimeter Institute.  I would also like to thank the Theory Group of Imperial College where part of this work was carried out.


\end{document}